\documentclass[pra,aps,multicol,epsfig,twocolumn]{revtex4}
\usepackage{graphicx}
\usepackage{amsmath}
\usepackage{amsfonts}
\usepackage{amssymb}
\usepackage{youngtab}
\newcommand{\beq}{\begin{equation}}
\newcommand{\eeq}{\end{equation}}

\begin{document}
\title{ Reduced density matrix of permutational invariant many-body systems}

\author{Mario Salerno$^1$ and Vladislav Popkov$^2$}
\affiliation{
$^1$ Dipartimento di Fisica \textquotedblleft E.R. Caianiello" and Istituto
Nazionale di Fisica Nucleare (INFN), Gruppo Collegato di Salerno,  Universit\'{a} di Salerno, via Ponte don Melillo,  I-84084
Fisciano (SA), Italy \\
$^2$Dipartimento di Fisica \textquotedblleft E.R. Caianiello" and Consorzio Nazionale per le Scienze Fisiche della Materia (CNISM),
Universit\'{a} di Salerno, via Ponte don Melillo,  I-84084
Fisciano (SA), Italy}
\thanks{E-mail: salerno@sa.infn.it}

\begin{abstract}
We consider density matrices which are sums of projectors on
states spanning irreducible representations of the permutation
group of $L$ sites (eigenstates of  permutational invariant
quantum system with $L$ sites) and construct reduced density
matrix $\rho_{n}$ for blocks of size $n<L$ by tracing out $L-n$
sites, viewed  as environment. Explicit analytic expressions of the
elements of $\rho_{n}$ are given in the natural basis and the
corresponding spectrum is derived. Results apply to
all quantum many-body systems with permutational symmetry for
which the mean field theory is  exact.
\end{abstract}

\date{\today}
\maketitle

\section{Introduction}

Reduced density matrices contain complete information about open quantum
i.e. quantum systems in contact with an environment such as a thermal
bath or a larger system of which the original system  constitute a part
(subsystem). In many cases one is interested in the spectrum of the reduced
density matrix (RDM) because it reveals intrinsic properties (sometimes universal)
of the subsystem. By definition,  the spectrum of the RDM is real and
nonnegative with all eigenvalues summing up to 1. The relative
importance of a state of the subsystem can then be measured by the
weight the corresponding eigenvalue has in the RDM spectrum. Thus, for instance,
the fact that the eigenvalues $\lambda_i$ of the RDM for a one dimensional quantum
interacting subsystems decay exponentially with $i$ implies that the properties of
the subsystems are determined by only a few states.
This property is crucial for the success of the density-matrix  renormalization group
(DMRG) method \cite{White92} in one dimension. In two dimensions this property is lost
\cite{PeschelChung} and the DMRG method fails.

For a subsystem consisting of $n$ sites (or $n$ q-bits) the  RDM is
of rank $2^{n}$ so that for large $n$ the calculation of the
spectrum becomes a problem of exponential difficulty. While the spectrum of the
full RDM for subsystems with a small number of sites (e.g. $n\leq6$) has been
calculated \cite{NienhuisCalabrese2009}, the full RDM  for arbitrary $n$, to our knowledge,
is exactly known only for the very special case of systems of free fermions (see e.g.
\cite{PeschelFreeFermionRDM}).

The aim of the present paper is to analytically calculate the elements of the RDM of
permutationally invariant quantum systems  of arbitrary size $L$,  for arbitrary
permutational symmetry of the state of the system (labeled by an integer  number $0<r<L/2$) and
for arbitrary sizes $n$ (number of q-bits) of the subsystem. We remark that the
invariance under the permutational group physically
implies that the interactions among sites have infinite range. From this point of view our
results may apply to all quantum systems for which mean field theory
becomes exact. As an example we consider a system of Heisenberg
spins $1/2$ on a full graph consisting of $L$ sites, with fixed value of
magnetization $S_{z}=L/2-N$. For this system  we calculate the RDM for a
subsystem of arbitrary $n\geq 1$ $\ $\ sites for arbitrary $L,N$.

The plan of the paper is the following.  The formulation of the main problem and the
basic definitions are given Section 2. In Section  3 we discuss general properties
of the elements of the RDM while in Section 4 we use symmetry properties of the system to
decompose  the RDM into block diagonal form. The main results of the paper are presented
in Sect. 5  in the form of a theorem giving  the analytical expressions of the RDM elements
for arbitrary $L,n,r$. For simplicity, we provide a proof of this theorem only  in the thermodynamic
limit. In Sec. 6 we characterize the spectrum of the RDM and discuss some of its main properties.
Finally, in the last section we briefly summarize the main results of the paper.

\section{Model equation and main definitions}

Consider a permutationally invariant system of $L$ spins $1/2$  on
a complete graph with fixed total magnetization $S_{z}=L/2-N$ and
described by the Hamiltonian
\begin{equation}
H=-\frac{J}{2L}\left( \mathbf{S}^{2}-\frac{L}{2}\left( \frac{L}{2}+1\right)
\right) +hS_{z}  \label{Hamiltonian}
\end{equation}
Here $\mathbf{S}\equiv(S_{x},S_{y},S_{z}),\;S_{\alpha}=\frac{1}{2}\sum
_{i=1}^{L}\sigma_{i}^{\alpha}$,  with $\sigma_{i}^{\alpha}$ Pauli
matrices acting on the factorized $\prod\limits_{1}^{L}\otimes C_{2}$ space.
This Hamiltonian is invariant under the action of the symmetric group $\mathbf{S}_{L}$ and
conserves the total spin polarization $S_{z}$, $[H,S_{z}]=0$. A complete set of eigenstates
of $H$ are states $|\Psi_{L,N,r}\rangle$ associated to filled Young Tableau (YT)\ of type
$\{L-r,r\}_{(N)}$ (see \cite{Mario94}),
\begin{eqnarray}
&&H|\Psi _{L,N,r}\rangle =E_{L,N,r}|\Psi _{L,N,r}\rangle , \\
&&E_{L,N,r}=\frac 1 2 \left( \frac{J r}{L}(L-r+1)+h(L-2N)\right) , \\
&&S_{z}|\Psi _{L,N,r}\rangle =\left( \frac{L}{2}-N\right) |\Psi
_{L,N,r}\rangle.
\end{eqnarray}
Here $N=0,1,...,L$ determines possible values of the spin polarization and
$r$ takes values $r=0,1,...,\max (N,L-N)$ (the explicit form of the state $|\Psi
_{L,N,r}\rangle $ is given below in Eq(\ref{YoungTableau_example})).
The degeneracies of the eigenvalues $E_{L,N,r}$ are given by the dimension of the
corresponding YTs,
\begin{equation}
\deg _{L,r}=\binom{L}{r}-\binom{L}{r-1}.
\end{equation}

{\bf \noindent \textit{Definition 1}} Consider a set of vectors $|\Psi_{u}\rangle$, $%
u=1,...\deg_{L,r}$, forming an orthonormal basis in the eigenspace of $H$
with eigenvalue $E_{L,N,r}$. We define the density matrix of the whole
system as
\begin{equation}
\sigma_{L,N,r}=\frac{1}{\deg_{L,r}}\sum_{u=1}^{\deg_{L,r}}|\Psi_{u}\rangle%
\langle\Psi_{u}|.  \label{thermal_density_matrix}
\end{equation}

\noindent $\sigma_{L,N,r}$ possess the following properties:

\noindent $\mathbf {i)}$
The matrix $\sigma _{L,N,r}$ has eigenvalues $\lambda _{1}=\lambda
_{2}=...=\lambda _{\deg _{L,r}}=(\deg _{L,r})^{-1}$,
 with remaining $2^{L}-\deg_{L,r}$ eigenvalues
all equal to zero. This follows from the fact that each vector
$|\Psi _{u}\rangle $ is an eigenvector of $\sigma_{L,N,r}$
with eigenvalue $\frac{1}{\deg _{L,r}}$. Since the spectrum of %
$\sigma_{L,N,r}$ is real and nonnegative with all eigenvalues
summing up to $1$, the remaining $2^{L}-\deg _{L,r}$ eigenvalues
must vanish.

\noindent $\mathbf {ii)}$ Matrix $\sigma_{L,N,r}$ satisfies:
$(\sigma_{L,N,r})^{2}=\frac{1}{\deg_{L,r}}\sigma_{L,N,r}$. This
follows from the definition (\ref{thermal_density_matrix}) and the
orthonormality condition $\langle\Psi_{w}|\Psi_{u}\rangle=\delta_{uw}$.

\noindent $\mathbf{ iii)}$  Introduce the operator $P_{ij}$, permuting subspaces $i$ and
$j$ of the  Hilbert space $\prod\limits_{1}^{L} \otimes C_{2}$ on which the matrix $\sigma_{L,N,r}$ acts.
Then $[\sigma,P_{ij}]=0$ for any $i,j$.

\noindent Proof.

\noindent Let us consider
\begin{eqnarray}
P_{ij}\sigma_{L,N,r} P_{ij} & = & \frac{1}{\deg_{L,r}}\sum_{u=1}^{%
\deg_{L,r}}P_{ij}|\Psi_{u}\rangle\langle\Psi_{u}|P_{ij} \nonumber \\ & = & \frac{1}{\deg_{L,r}}
\sum_{u=1}^{\deg_{L,r}}|\Psi_{u} ^{\prime}\rangle\langle\Psi_{u}^{\prime}|.
\end{eqnarray}
The vectors $|\Psi_{u}^{\prime}\rangle=P_{ij}|\Psi_{u}\rangle$ form an
orthonormal basis. Indeed, $\langle\Psi_{w}^{\prime}|\Psi_{u}^{\prime}%
\rangle=\langle\Psi
_{w}|P_{ij}^{T}P_{ij}|\Psi_{u}\rangle=\langle\Psi_{w}|\Psi_{u}\rangle
=\delta_{uw}$, because $P_{ij}^{T}=P_{ij}$, and $(P_{ij})^{2}=I$. Now, the
sum $\sum_{u=1}^{\deg_{L,r}}|\Psi_{u}\rangle\langle\Psi_{u}|=I_{\deg_{L,r}}$
is a unity operator in a subspace of dimension $\deg_{L,r}$, and therefore
it does not depend on the choice of the basis. Note that vector $|\Psi_{u}^{\prime}\rangle$
belongs to the same subspace as $|\Psi _{u}\rangle$, because permutation $P_{ij}$ only results
in different enumeration. Consequently,
\begin{equation}
P_{ij}\rho P_{ij}=\rho\text{, \ \ \ or \ }[\rho,P_{ij}]=0.
\label{Prho-rhoP}
\end{equation}
The latter property implies that in Eq. (\ref{thermal_density_matrix}) the
sum over the orthogonalized set of basis vector in (\ref{thermal_density_matrix}) can be
replaced by the  symmetrization of  the density matrix directly, namely $\sigma_{L,N,r}=
\frac{1}{L!}\sum_{P}|\Psi_{12...L}\rangle\langle\Psi_{12...L}|$, where the sum is
over all $L!$ permutations of indexes $1,2,...L$, and $|\Psi_{12...L}\rangle$
is some unit eigenvector of $H$ with eigenvalue $E_{L,N,r}$. In particular,
it is convenient to choose $|\Psi_{12...L}\rangle\equiv|\Psi_{L,N,r}\rangle$,
\begin{equation}
\sigma_{L,N,r}=\frac{1}{L!}\sum_{P}|\Psi_{L,N,r}\rangle\langle\Psi_{L,N,r}|.
\label{sigma_as_SumOfPermutaitons}
\end{equation}
It is evident that such a sum is invariant with respect to
permutations and that $\sigma_{L,N,r}$ is properly normalized:
$Tr\sigma_{L,N,r}=1$.

{\bf \noindent \textit{Definition 2}} The Reduced Density Matrix (RDM) of a
subsystem of $n$ sites is defined by tracing out $L-n$ degrees of
freedom from the density matrix of the whole system:
\begin{equation}
\rho_{(n)}=Tr_{L-n}\sigma_{L,N,r}.  \label{RDM}
\end{equation}
Due to the properties (\ref{Prho-rhoP}) and (\ref{RDM}), $\rho_{(n)}$ does
not depend on the particular choice of the $n$ sites, and satisfies the
property (\ref{Prho-rhoP}) in its subspace (we omit the explicit
dependence of $\rho_{(n)}$ on $L,N,r$ for brevity of notations).

\section{Properties of the RDM}

The RDM can be calculated in the natural basis  by using its definition in terms of
observables:
$\langle\hat{f}\rangle=Tr(\rho_{(n)}\hat{f})$ where $\hat{f}$ is a physical
operator acting on the Hilbert space of the $2^{n}\times2^{n}$ subsystem.
The knowledge of the full set of observables determines the RDM uniquely.
Indeed, if we introduce the natural basis in the Hilbert space of the
subsystem, $\prod\limits_{k=1}^{n}\otimes C_{2}$, the elements of the
RDM in this basis are
\begin{equation}
\rho_{j_{1}j_{2}...j_{n}}^{i_{1}i_{2}...i_{n}}=\left\langle \hat{e}%
_{j_{1}j_{2}...j_{n}}^{i_{1}i_{2}...i_{n}}\right\rangle =Tr\left( \rho _{(n)}%
\hat{e}_{j_{1}j_{2}...j_{n}}^{i_{1}i_{2}...i_{n}}\right) ,
\end{equation}
with $\hat{e}_{j_{1}j_{2}...j_{n}}^{i_{1}i_{2}...i_{n}} =
\prod\limits_{k=1} ^{n}\otimes \hat{e}_{j_{k}}^{i_{k}}$ and $\hat{e}_{j}^{i}$
a $2\times2$ matrix with elements $\left( \hat{e}%
_{j}^{i}\right) _{kl}=\delta_{ik}\delta_{jl}$. The matrix $\hat{e}%
_{j_{1}j_{2}...j_{n}}^{i_{1}i_{2}...i_{n}}$ has only one nonzero element,
equal to $1$, at the crossing of the row
$2^{n-1}i_{1}+2^{n-2}i_{2}+...+i_{n}+1$ and the column $%
2^{n-1}j_{1}+2^{n-2}j_{2}+...+j_{n}+1$ (all indices $i,j$ take binary values
$i_{k}=0,1$ and $j_{k}=0,1$). To determine all the RDM elements one must  find
a complete set of observables and compute the averages $\left\langle \hat{e}_{j_{1}j_{2}...j_{n}}
^{i_{1}i_{2}...i_{n}}\right\rangle$. Note that a generic property of the
RDM elements, which follows directly from (\ref{Prho-rhoP}), is that any
permutation between pairs of indices $(i_{m},j_{m})$ and $(i_{k},j_{k})$
does not change its value, e.g.
\begin{equation}
\rho_{j_{1}j_{2}...j_{n}}^{i_{1}i_{2}...i_{n}}=%
\rho_{j_{2}j_{1}...j_{n}}^{i_{2}i_{1}...i_{n}}=%
\rho_{j_{n}j_{1}...j_{2}}^{i_{n}i_{1}...i_{2}}=...= \rho_{j_{n}j_{n-1}...j_{1}}^{i_{n}i_{n-1}...i_{1}}.
\label{RDM_elements_Permute}
\end{equation}
Another property of the RDM follows from the $S_{z}$ invariance
\begin{equation}
\rho_{j_{1}j_{2}...j_{n}}^{i_{1}i_{2}...i_{n}} = 0,  \text{\qquad if }i_{1}
+i_{2}+...+i_{n}\neq j_{1}+j_{2}+...+j_{n}.  \label{RDM_elements_IceRule}
\end{equation}
Thus, for instance, the RDM for $n=2$ has only three different nonzero \
elements, $\rho_{00}^{00},\rho_{01}^{10}=\rho_{10}^{01} $and $\rho_{11}^{11}$. It is convenient
to introduce the operators
\begin{eqnarray}
&& \hat{e}_{0}^{1}=
\begin{pmatrix}
0 & 0 \\
1 & 0%
\end{pmatrix}
\equiv \sigma^{-},\;\;\;\;\; \hat{e}_{1}^{0}=%
\begin{pmatrix}
0 & 1 \\
0 & 0%
\end{pmatrix}
\equiv\sigma^{+},  \notag \\
&& \hat{e}_{0}^{0}=%
\begin{pmatrix}
1 & 0 \\
0 & 0%
\end{pmatrix}
\equiv\hat{p}\;,\;\;\;\;\; \hat{e}_{1}^{1}=%
\begin{pmatrix}
0 & 0 \\
0 & 1%
\end{pmatrix}
\equiv\hat{h}\;.  \notag
\end{eqnarray}
If we represent a site spin up  with the vector $\binom{1}{0}$ and
a site spin down with the
 vector $\binom{0}{1}$ then $\hat{p}_{k}$ and $\hat{h}_{k}$ are
spin up and spin down number operators on site $k$, while $\sigma
^{-},\sigma^{+}$ represent spin lowering and rising operators, respectively.
Thus, for instance, the observable $\langle\hat{p}_{1}\hat{p}_{2}\hat{h}_{3}%
\hat {h}_{4}...\hat{h}_{n}\rangle=$ $\rho_{0011...1}^{0011...1}$ gives the
probability to find spins down at sites $3,4,...n$, and spins up at sites $%
1,2$, while the observable $\langle\sigma_{1}^{+}\sigma_{2}^{+}%
\sigma_{3}^{-}\sigma_{4}^{-}\hat{h}_{5}...\hat{h}_{n}\rangle=%
\rho_{11001...1}^{00111...1}$ gives the probability to find spins down at
sites $5,6,...n$, spin lowering at sites $3,4$ and spin rising at sites $1,2$%
. Note that the latter operator conserves the total spin polarization
since the number of lowering and rising operators is the same. Also note that the
correlation functions with
a non conserved polarization vanish,
e.g. $$\langle \sigma_{1}^{+}
\sigma_{2}^{+} \sigma_{3}^{-} \hat{h}_{4} \hat{h}_{5}...\hat{h}
_{n}\rangle = \rho_{11011...1}^{00111...1} = 0,$$ in accordance with (\ref
{RDM_elements_IceRule}).

\section{Block diagonal form of the RDM}

\begin{figure}[h]
\centerline{\scalebox{.4}{\includegraphics{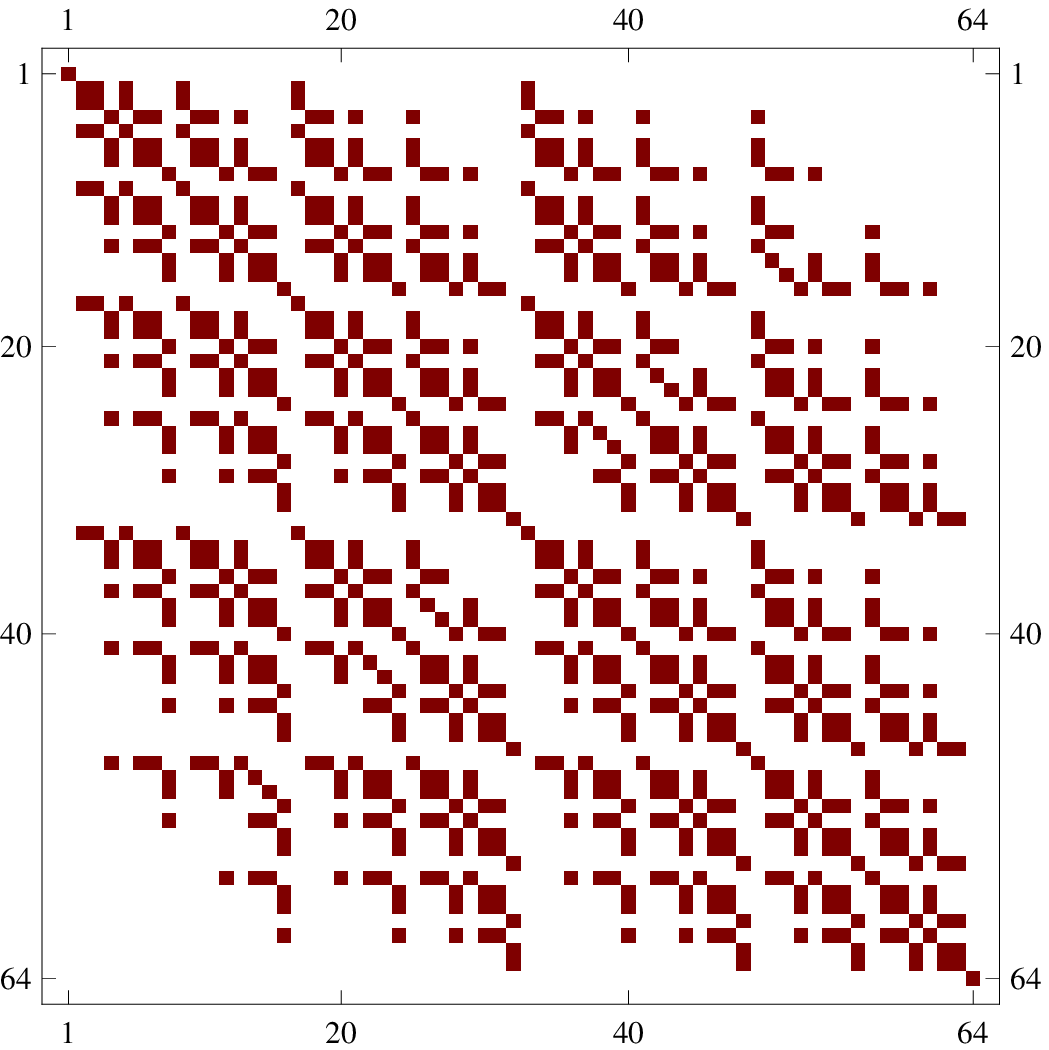}}
\scalebox{.4}{\includegraphics{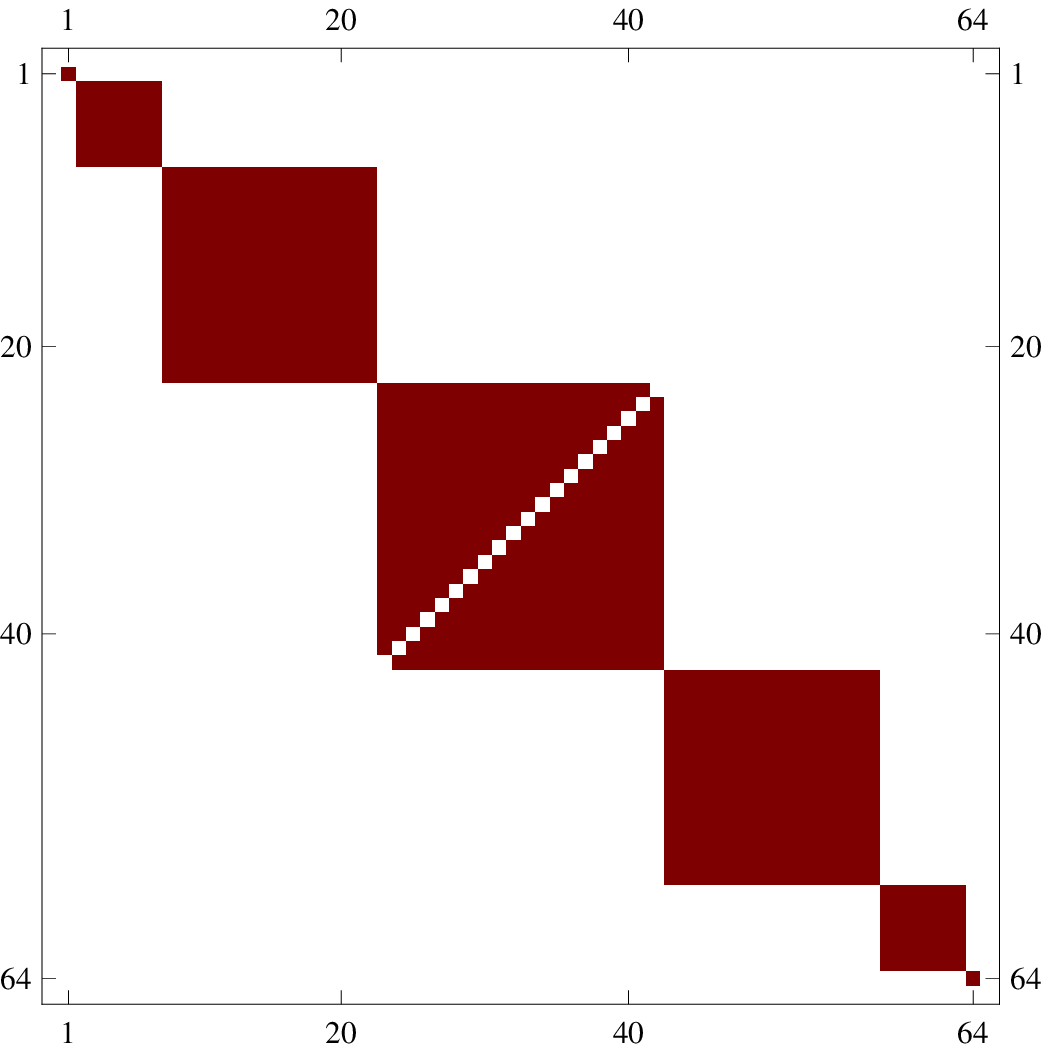}} } \caption{Left
panel. Reduced Density Matrix $\rho_{(6)}$ in the natural basis.
Parameters are $L=12,N=6,r=3$. Black boxes denote non zero
elements. Right panel. The same matrix of the  left panel after
the following chain of permutations of rows and columns have been
applied to it:
$P_{22,25}P_{21,34}P_{20,35}P_{16,37}P_{15,41}P_{14,33}P_{12,21}P_{8,49}P_{7,33}P_{6,17}P_{4,9}$.
($P_{i,j}$ denote the operator which exchange first columns $i$
and $j$ and then rows $i$ and $j$). Black boxes denote non zero
elements. Block diagonal structure associated with values of
$k=0,1,...,6$ is evident. The single element present in blocks
$k=0, 6,$ is  $1/924$. Elements values  inside
other k-blocks are given in Fig. \ref{Fig_n6_twoblocks}%
. }
\label{Fig_rho6_sparse}
\end{figure}

One can take advantage of the $S_z$  invariance (e.g.  Eq. \ref{RDM_elements_IceRule})
to block diagonalize the RDM into independent blocks $B_{k}$ of fixed polarization
$k=i_{1}+i_{2}+...+i_{n} =j_{1}+j_{2}+...+j_{n}$ (here $k$ gives the
number of spin up present in the subsystem). Note (see Fig.\ref{Fig_rho6_sparse})
that the block diagonal form in the natural basis  becomes evident
after a number of permutations of rows and columns of the RDM
have been  performed. Also notice that the $n+1$ blocks correspond to the values $s_z=(n
- 2 k)/2$, $k=0,1,..., n$ the subsystem polarization can assume.
The dimension of a block $B_{k}$
coincides with the number of possible configurations that $k$ spin
up can assume on $n$ sites, e.g. $\dim B_{k}=\binom{n}{k}$. One
can easily check that the sum of the dimensions of all blocks
gives the full RDM dimension, i.e.  $\sum_{k}\dim B_{k}=2^{n}$.

Blocks $B_{k}$ consist of elements $e_{j_{1}}^{i_{1}}\otimes
e_{j_{2}}^{i_{2}}\otimes e_{j_{3}}^{i_{3}}\otimes...\otimes
e_{j_{n}}^{i_{n}}$ \ of the original
matrix, with $\sum\limits_{1}^{n}i_{p}=\sum\limits_{1}^{n}j_{p}=k$ and $%
i_{p}=0,1$, $j_{p}=0,1$. In its turn, all elements
$e_{j_{1}}^{i_{1}}\otimes e_{j_{2}}^{i_{2}}\otimes
e_{j_{3}}^{i_{3}}\otimes...\otimes e_{j_{n}}^{i_{n}} $ of the
block $B_{k}$ can be further block diagonalized according to the
irreducible representations of the permutation group $\bf S_n$ of the
subsystem (see also section 6 below). In the natural basis, this
block diagonalization is done according to the number of pairs of type
$(e_{1}^{0}\otimes e_{0}^{1})$ present in the elements. In
the following we denote by $G_{Z}$ the part of the block
associated to elements with $Z$ pairs $(e_{1}^{0}\otimes
e_{0}^{1})$ in it. The sub-block $G_{0}$ of the block $B_{k}$ is
formed by the elements containing $k$ terms $e_{1}^{1}$ and
($n-k$) terms $e_{0}^{0}$ in the product, i.e.
$e_{1}^{1}\otimes...e_{1}^{1}\otimes e_{0}^{0}\otimes...\otimes
e_{0}^{0}$ and all permutations. All such
elements lie on the diagonal, and vice versa, each diagonal element of $%
B_{k} $ belongs to $G_{0}$. Consequently, the sub-block $G_{0}$ consists of $%
\binom{n}{k}$ elements.

The number of elements, $\deg G_{1}(k)$, in the sub-block $G_{1}$ is equal to
the number of elements of the type $e_{1}^{0}\otimes e_{0}^{1}\otimes
e_{i_{1} }^{i_{1}}\otimes e_{i_{2}}^{i_{2}}\otimes...\otimes
e_{i_{n-2}}^{i_{n-2}}$, such that $1+0+i_{1}+i_{2}+...+i_{n-2}=k$. Using
elementary combinatorics we obtain:
\begin{equation}
\deg G_{1}(k)=\binom{2}{1}\binom{n}{2}\binom{n-2}{k-1}.
\end{equation}
\begin{figure}[h]
\centerline{\scalebox{.9}{\includegraphics{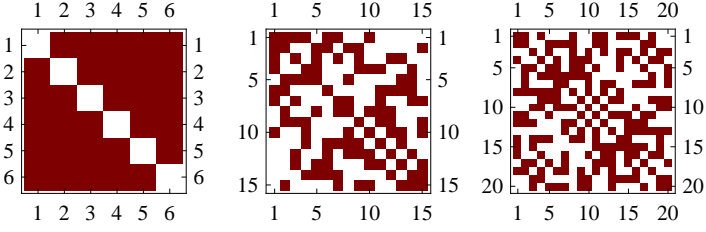}}}
\caption{Blocks $k=1, 5$ (left), $k=2,4$ (center) and $k=3$
(right) appearing in the block diagonal form of $\rho_{(6)}$ given
in the right panel of Fig.\ref{Fig_rho6_sparse}. White boxes in
the left panel denote the element $1/154$ while black boxes denote
the element $1/924$. In the center panel white boxes denote the
element $5/1848$ while black boxes, except for the diagonal
elements which are equal to $5/308$, denote the element $-1/924$.
In the right panel white boxes denote the element $5/1386$ except
for the ones along the anti-diagonal which are all zero, while the
black boxes denote the element $-1/693$, except for the ones along
the diagonal which are all $5/231$.} \label{Fig_n6_twoblocks}
\end{figure}
Analogous calculations for arbitrary sub-block $G_{Z}$ yields
\begin{equation}
\deg G_{Z}(k)=\binom{2Z}{Z}\binom{n}{2Z}\binom{n-2Z}{k-Z}.  \label{deg_Gz(k)}
\end{equation}
From the restriction $\sum_{1}^{n}i_{p}=\sum_{1}^{n}j_{p}=k$ we deduce that
the block $B_{k}$ contains non-empty parts $G_{0}, G_{1},...,  G_{\min(k,n-k)}$,
leading to the following decomposition:
\begin{equation}
B_{k}=\bigcup\limits_{Z=0}^{\min(k,n-k)}G_{Z}.  \label{Bk_decomposition}
\end{equation}
Indeed, the normalization condition following from (\ref{Bk_decomposition}),
gives
\begin{equation}
\sum_{Z=0}^{\min(k,n-k)}\deg G_{Z}(k)=\binom{n}{k}^{2}.
\end{equation}

It is important to note that \textit{all elements of} $G_{Z}$ \textit{are equal}.
This is a direct consequence of the property in Eq. (\ref{RDM_elements_Permute}). A graphical
representation  of the $B_k$ block for a particular choice of $L,N,r$, is given in
Fig.\ref{Fig_n6_twoblocks}.

\section{Analytical expression for RDM elements}

The main result of the paper is provided by the following theorem.

{\bf \textit{Theorem}} Elements of RDM $g_{Z}$, belonging to a sub-block $G_{Z}$
of a block $B_{k}$ of the reduced density matrix (\ref{RDM}) are given, for arbitrary
$L,N,r,n$,  by
\begin{equation}
g_{Z}=\frac{\binom{L-n}{N-k}}{\binom{L}{N}}\frac{\sum_{m=0}^{Z}\left(
-1\right) ^{m}\binom{N-r}{Z-m}\binom{L-N-r}{Z-m}\binom{r}{m}}{\binom{N}{Z}%
\binom{L-N}{Z}}.  \label{g_z}
\end{equation}
This expression defines completely all elements of the RDM in the natural
basis. In practice, to find the element $P,Q$ of the RDM, $\left( \rho
_{(n)}\right) _{PQ}$, we have to take the binary representation of numbers
$P-1 $ and $Q-1$ (which gives us the sets $\{i_{p}\}$ and $\{j_{p}\}$
respectively), find the corresponding number $Z$, and use (\ref{g_z}).

The proof of this theorem will be given in the thermodynamic limit, $L\rightarrow
\infty,\frac{N}{L}=p,\frac{r}{L}=\mu$, in which the expression (\ref{g_z})
simplifies to
\begin{equation}
g_{Z}=p^{n-k}(1-p)^{k}\eta^{Z},  \label{Gz_elements_thermodynamic_limit}
\end{equation}
with
\begin{equation}
\eta=\frac{(p-\mu)(1-p-\mu)}{p(1-p)}.  \label{eta}
\end{equation}

We shall prove Eq. (\ref{Gz_elements_thermodynamic_limit}) first for the case $r=0$ corresponding to
the symmetric ground states in (\ref{GS}) and then for the case $r>0$ corresponding to excited states.

\subsection{Case $r=0$.}

If $r=0$, the Young Tableau $Y_{L,N,r}$ is nondegenerate and the state of
the initial system is pure: $\rho=|\Psi_{L,N}\rangle\langle\Psi_{L,N}|$
with $|\Psi_{L,N}\rangle$ the symmetric state
\begin{equation}
|\Psi_{L,N}\rangle=\binom{L}{N}^{-1/2}\sum_{P}|\underbrace{\uparrow
\uparrow...\uparrow}_{N}\underbrace{\downarrow\downarrow...\downarrow\rangle
}_{L-N}  \label{GS}
\end{equation}
where the sum is over all possible permutations, and the prefactor takes
care of normalization. Since all $L$ sites are equivalent due to
permutational invariance, any choice of $n$ sites $i_{1},i_{2},...,i_{n}$
within $L$ sites gives the same RDM, which we denote by
$\rho_{(n)}^{L,N,0}=Tr_{L-n}\rho$. It has been shown in \cite{Entanglement_Heisenberg}
that $\rho_{(n)}^{L,N,0}$ takes form
\begin{equation*}
\rho_{(n)}^{L,N,0}=\sum_{k=0}^{n}\frac{\binom{L-n}{N-k}}{\binom{L}{N}}
|\Psi_{n,k}\rangle\langle\Psi_{n,k}|.
\end{equation*}
In the natural basis the matrix elements of RDM are given by the above
discussed values of observables. Using (\ref{GS}), one explicitly computes
all RDM elements as%
\begin{equation}
\left( \rho_{(n)}^{L,N,0}\right) _{Q}^{P}=\left( \rho_{(n)}^{L,N,0}\right)
_{i^{\prime}j^{\prime}...m^{\prime}}^{ij...m}=\delta_{i^{\prime}+j^{\prime
}...+m^{\prime}}^{i+j+..+.m}\frac{\binom{L-n}{N-w}}{\binom{L}{N}}
\label{RDM_r=0}
\end{equation}
with $w=i+j+...+m$   (the sets $ij...m$ and $i^{\prime}j^{\prime}...m^{\prime}$ are binary
reprentation of numbers $P-1,Q-1$). So, all the elements of a given block $B_{k}$
are equal. This property does not hold in the general case $r\geq0$. \ Note
that in the thermodynamic limit $\eta=1$, and $\binom{L-n}{N-k}/\binom{L}{N}%
\rightarrow p^{n-k}(1-p)^{k}$.

\subsection{General case $r>0$.}
To calculate the RDM, we shall use the representation (\ref%
{sigma_as_SumOfPermutaitons}) for the density matrix of the whole system $%
\sigma$, rewriting it in the following form
\begin{eqnarray}
\rho_{(n)} & = & Tr_{L-n}\left\{ \frac{1}{L!}\sum_{P}|\Psi_{L,N,r}\rangle\langle
\Psi_{L,N,r}| \right\}  \nonumber \\
&= & Tr_{L-n}\left\{\frac{1}{n!}\frac{1}{(L-n)!}\frac{1}{\binom{L}{n}}%
\sum_{P_{(n)}}\sum_{P_{(L-n)}} \nonumber \right. \\ && \left. \sum_{i_{1}\neq i_{2}\neq...\neq
i_{n}}|\Psi_{L,N,r}\rangle\langle\Psi_{L,N,r}|\right\}.
\label{PermutationSplittingInSteps}
\end{eqnarray}
\begin{figure}[h]
\beq
{\large \Yvcentermath1\young(1111110,0000)} \nonumber
\eeq
\vskip .5cm
\caption{Graphical representation of a filled Young Tableau $\{L-r,r\}_{N}$
with $L=11,r=4,N=6$. The $1$ inside a box denote a spin up (particle) while the
$0$ denote  a spin down (hole).}
\label{Fig_YT}
\end{figure}

Note that the  $L!$ permutations can be done in three steps: first, choose at random $n$
sites $i_{1}\neq i_{2}\neq...\neq i_{n}$ among the $L$ sites. There are $\binom{L%
}{n}$ such choices. Then, permute the chosen $n$ sites, the total number of
such permutations being $n!$. Finally, permute the remaining $L-n$ sites,
the total number of such permutations being $(L-n)!$. The latter step (c)
under the trace operation is irrelevant because these degrees of freedom
will be traced out. The operation permuting $n$ sites commutes with the
trace operation since $Tr_{L-n}$ does not touch the respective subset of $n$
sites. Consequently, (\ref{PermutationSplittingInSteps}) can be rewritten as
\begin{equation}
\rho_{(n)}=\frac{1}{n!}\sum_{P_{(n)}}Tr_{L-n}\frac{1}{\binom{L}{n}}
\sum_{i_{1}\neq i_{2}\neq...\neq i_{n}}|\Psi_{L,N,r}\rangle\langle
\Psi_{L,N,r}| \,.  \label{RDM_operational}
\end{equation}

The filled Young tableau $\{L-r,r\}_{N}$ consists of the antisymmetric part
containing $2r$ sites and $r$ spin up, and the symmetric part containing
the remaining $L-2r$ sites and $N-r$ spin up, see Fig. \ref{Fig_YT}
(in the following  we adopt  an equivalent terminology  appropriate for non spin systems by
identifying spins up with particles and spins down with holes). The wave function $|\Psi_{L,N,r}\rangle$,
corresponding to this Young tableau factorizes into symmetric and antisymmetric parts
\begin{equation}
\Psi =|\phi _{12}\rangle \otimes |\phi _{34}\rangle \otimes ...\otimes |\phi
_{2r-1,2r}\rangle \otimes |\Psi _{L-2r,N-r}\rangle _{2r+1,2r+2,..L}
\label{YoungTableau_example}
\end{equation}%
with the antisymmetric part consisting of the first $r$ factors of the type
\begin{equation}
|\phi _{12}\rangle =\frac{1}{\sqrt{2}}(|10\rangle _{12}-|01\rangle _{12})
\label{asym_wavefunction}
\end{equation}
and with the symmetric part, $|\Psi _{L-2r,N-r}\rangle $,  given by (\ref
{GS}). A general property of factorized states implies that if the
global wave function is factorized, $|\Phi \rangle =|\psi \rangle _{I}|\phi
\rangle _{II}$ and out of $n$ sites of the subsystem, $n_{1}$ sites belong
to subset $I$, and the remaining $n_{2}=n-n_{1}$ sites belong to the subset $%
II$, then the reduced density matrix factorizes as well:
\begin{equation}
\rho _{(n)}=\rho _{(n_{1})}^{I}\otimes \rho _{(n_{2})}^{II}.
\label{density_matrix_product}
\end{equation}

To do the averaging, we note that among total number of choices $\binom{L}{n}
$ there are (a) $\binom{L-2r}{n}$ possibilities to choose $n$ sites inside
the symmetric part of the tableau, containing $N-r$ particles, (b) $2r\binom{%
L-2r}{n-1}$ possibilities to choose $n-1$ sites inside the symmetric part of
the tableau and one site in the antisymmetric part (c) $\binom{2r}{2}\binom{%
L-2r}{n-2}$ possibilities to choose $n-2$ sites inside the symmetric part of
the tableau and two sites in the antisymmetric part and so on. The contributions
given by (a) and (b) to the right hand side of (\ref
{RDM_operational}) for $\rho _{(n)}^{L,N,r}$ are given, according to (\ref
{density_matrix_product}), by
\beq
\left\langle \binom{F}{n}\rho _{(n)}^{F,M,0}+2r\binom{F}{n-1}\rho
_{(n-1)}^{F,M,0}\otimes \rho _{\frac{1}{2}}\right\rangle,
\eeq
with $F=L-2r$, $M=N-r$ and with $\rho _{\frac{1}{2}}=I/2$  the
density matrix corresponding to a single site in the antisymmetric part of
the tableau. Brackets $\langle .\rangle =\frac{1}{n!}\sum_{P}$ denote the
average with respect to  permutations of $n$ elements.The contribution \ due to (c) to
(\ref{RDM_operational}) splits into two parts since the $\binom{2r}{2}$ possibilities
to choose two sites in the antisymmetric part of the tableau consist of
$4 \binom{r}{2}$ choices with two sites into different columns and
the remaining $r$ choices with  both sites belonging to a same column. For the former
choice, the corresponding density matrix is $\rho _{(n-2)}^{F,M,0}\otimes
\rho _{\frac{1}{2}}\otimes \rho _{\frac{1}{2}}$, while for the latter case
is given by $\rho _{(n-2)}^{F,M,0}\otimes \rho _{asymm}$, with
\begin{equation}
\rho _{asymm}=|\frac{1}{\sqrt{2}}(10-01)\rangle \langle \frac{1}{\sqrt{2}}%
(10-01)|=\frac{1}{2}%
\begin{pmatrix}
0 & 0 & 0 & 0 \\
0 & 1 & -1 & 0 \\
0 & -1 & 1 & 0 \\
0 & 0 & 0 & 0%
\end{pmatrix}.  \label{rho_asymm}
\end{equation}%
Proceeding in the same manner for arbitrary partitions of $Z$ sites in the
antisymmetric part of the tableau and $n-Z$ sites in the symmetric part, we
get
\begin{eqnarray}
&&
\rho _{(n)}^{L,N,r} \binom{L}{n} = \left \langle\sum_{Z=0}^{\min (2r,n)}
\binom{F}{n-Z} \rho_{(n-Z)}^{F,M,0} \sum_{i=0}^{[Z/2]}\binom{r}{i} \right.
\nonumber \\
&& \left. \left({\prod\limits_{1}^{i}}\otimes \rho_{asymm}\right) 2^{Z-2i}
\binom{r-i}{Z-2i}\left( {\displaystyle\prod\limits_{1}^{Z-2i}}\otimes
\rho_{\frac{1}{2}}\right) \right\rangle.
\label{thermal_density_matrix_decomposition1}
\end{eqnarray}
From this the general scheme  for the decomposition of the general RDM becomes evident. In
the above formula, the products $\prod\limits_{i}^{Q}$ with $Q<i$ are
discarded. The matrix elements $\rho _{(k)}^{F,M,0}$ are given by (\ref{RDM_r=0}).

For simplicity of presentation, we prove Eq. (\ref{Gz_elements_thermodynamic_limit})
for the case $Z=k$ and then outline the proof for arbitrary $Z$.

In the thermodynamic limit one can neglect the difference between factors like
$4\binom{r}{2}$ and $\binom{2r}{2}$ in Eq. (\ref{thermal_density_matrix_decomposition1}).
The latter then can be then rewritten in a simpler form as
\begin{align}
\binom{L}{n}\rho_{(n)}^{L,N,r} & =\left\langle \binom{F}{n}\rho
_{(n)}^{F,M,0}+\binom{2r}{1}\binom{F}{n-1}\rho_{(n-1)}^{F,M,0}\otimes \rho_{%
\frac{1}{2}}\right.
\label{thermal_density_matrix_thermodynamic_limit_decomposition} \\
& \left. +\binom{2r}{2}\binom{F}{n-2}\rho_{(n-2)}^{F,M,0}\otimes\rho _{\frac{%
1}{2}}\otimes\rho_{\frac{1}{2}}+...\right\rangle .  \notag
\end{align}
Note that one can omit all terms in (\ref%
{thermal_density_matrix_decomposition1}) containing $\rho_{asymm}$ since the
respective coefficients correspond to probabilities of finding two adjacent sites
in the asymmetric part of the YT
(proportional to $r$), which vanish in the thermodynamic limit, respect
to the total number of choices which is of order of $r^{2}$. A sub-block $%
G_{Z}$ of a block $k$ consists of all elements of the matrix $\rho_{(n)}$
having $Z$ pairs of $e_{1}^{0},e_{1}^{0}$ in its tensor representation, like
e.g. $\left( e_{1}^{0}\otimes e_{0}^{1}\right) ^{\otimes_{Z}}\otimes
e_{i_{1}}^{i_{1}}\otimes e_{i_{2}}^{i_{2}}\otimes...\otimes
e_{i_{n-2Z}}^{i_{n-2Z}}$, such that $Z+i_{1}+i_{2}+...+i_{n-2Z}=k$. The
total number of elements $g_{Z}\subset G_{Z}$ in $\rho_{(n)}^{L,N,r}$ is
equal to the number of distributions of $Z$ objects $e_{1}^{0}$, $Z$ objects
$e_{0}^{1}$, and $(k-Z)$ objects $e_{1}^{1}$ on $n$ places, given by
\beq
\deg G_{Z}=\frac{n!}{Z!Z!(k-Z)!(n-k-Z)!}
\eeq
(this is another way of writing (\ref{deg_Gz(k)})). Each term $W$ in the sum
(\ref{thermal_density_matrix_thermodynamic_limit_decomposition}) after
averaging will acquire the factor
\begin{equation}
\Gamma(W)=\frac{\deg G_{Z}(W)}{\deg G_{Z}}  \label{averaging_factor}
\end{equation}
where $\deg G_{Z}(W)$ is a total number of $g_{Z}$ elements in the term $W$,
provided all of them are equal. For instance, $\deg
G_{Z}(\rho_{(n)}^{F,M,0})=\deg G_{Z}$, \ $\deg
G_{Z}(\rho_{(n-m)}^{F,M,0}\otimes\left( \rho_{\frac{1}{2}}\right)
^{\otimes_{m}})=\binom{2Z}{Z}\binom{n-m}{2Z}\sum_{m_{1}=0}^{m}\binom{m}{m_{1}%
}\binom{n-2Z-m}{k-Z-m_{1}}$( the last formula is only true for $k=Z$,
otherwise elements constituting $G_{Z}(W)$ are not all equal). Restricting
to the case $k=Z$ and denoting $W_{m}=\rho_{(n-m)}^{F,M,0}\otimes\left(
\rho_{\frac{1}{2}}\right) ^{\otimes_{m}}$, we have
\begin{equation}
\Gamma(W_{m})=\Gamma_{m}=\frac{\binom{n-m}{2Z}}{\binom{n}{2Z}}.
\label{averaging_factor_m}
\end{equation}
It is worth to note that the element $g_{Z}\subset G_{Z}$ is simply given by
\begin{eqnarray}
\binom{L}{n}g_{Z}=\Gamma_{0}\binom{F}{n}g_{0}^{(n,k)}+\Gamma_{1}\binom{2r}{1}%
\binom{F}{n-1}\frac{g_{0}^{(n-1,k)}}{2} \nonumber + \\ \Gamma_{2}\binom{2r}{2}\binom {F}{n-2%
}\frac{g_{0}^{(n-2,k)}}{2^{2}}+ ...  \qquad\qquad\qquad\qquad \label{c_element_thermodynamic_limit}
\end{eqnarray}
with $q=1-p$ and $g_{0}^{(n,k)}=\frac{\binom{F-n}{M-k}}{\binom{F}{M}}%
\approx\left( \frac{p-\mu}{1-2\mu}\right) ^{^{n-k}}\left( \frac{q-\mu }{%
1-2\mu}\right) ^{k}$ is the element of a $\rho_{(n)}^{F,M,0}$ 
corresponding to a block with $k$ particles (the factors $\Gamma_{m}$ are due to
the averaging while the factors $\frac{1}{2^{m}}$ come from $\left( \rho_{\frac{1}{2}
}\right) ^{\otimes_{m}}$). Restricting to the case $k=Z$, and taking into
account
\beq
\frac{\binom{F}{n-m}}{\binom{L}{n}}  \approx\frac{n!}{(n-m)!}\frac
{(1-2\mu)^{n}}{F^{m}}, \;\;\;\;\;\;\;
\binom{2r}{m}  \approx\frac{(2\mu)^{m}}{m!}L^{m}\;,
\eeq
so that
\begin{equation*}
\frac{\binom{F}{n-m}}{\binom{L}{n}}\binom{2r}{m}\frac{1}{2^{m}}\approx
\binom{n}{m}\mu^{m}(1-2\mu)^{n-m},
\end{equation*}
we finally  obtain, using (\ref{c_element_thermodynamic_limit}), that
\begin{eqnarray}
g_{Z} & = &\sum_{m=0}^{n-2Z}\mu^{m}\binom{n-2Z}{m}(p-\mu)^{n-m-Z}(q-\mu
)^{Z} \nonumber \\  & = & (p-\mu)^{n-Z}(q-\mu)^{Z}\sum_{m=0}^{n-2Z}\frac{\mu^{m}}{(p-\mu)^{m}}
\binom{n-2Z}{m} \nonumber  \\
& = &(p-\mu)^{n-Z}(q-\mu)^{Z}\left( \frac{p}{p-\mu}\right)
^{n-2Z} \label{g_Z_thermodynamic_k=Z} \\ & = & p^{n-Z}q^{Z}\left( \frac{(p-\mu)(q-\mu)}{pq}\right)
^{Z} \nonumber \\ & = & p^{n-Z}q^{Z}\eta^{Z} = \eta^{Z}g_{0} \,, \nonumber
\end{eqnarray}
with $g_{0}$ the diagonal element in the same block $k=Z$. In the last
calculation we used the relation
$\frac{\binom{n}{m}\binom{n-m}{2Z}}{\binom {n}{n-Z}}=\binom{n-2Z}{m}$. This
proves formula (\ref{Gz_elements_thermodynamic_limit}) for the particular
case $k=Z$ and arbitrary $n$.

For arbitrary $k,Z,$ one proceeds in similar manner as for the case $k=Z$
case. Since the respective calculations are tedious and not particularly
illuminating,  we omit them  and give only the final result:
\begin{eqnarray}
g_{Z} & = & \sum_{m=0}^{n-2Z}\mu^{m}\sum_{i=\max(0,Z+k-n+m)}^{\min(m,k-Z)}(p-
\mu)^{n-m-k+i}  \nonumber \\
& & (q-\mu)^{k-i}\binom{k-Z}{i}\binom{n-k-Z}{m-i},
\end{eqnarray}
which, after some algebraic manipulation, can  be rewritten in the form
\begin{eqnarray}
g_{Z} & = & (p-\mu)^{n-k}(q-\mu)^{k}\sum_{j=0}^{n-k-Z}\left( \frac{\mu }{
p-\mu}\right) ^{j}\binom{n-k-Z}{j}  \nonumber\\
& & \times \sum_{i=0}^{k-Z}\left(
\frac{\mu}{q-\mu}\right) ^{i}\binom{k-Z}{i} = \eta^{Z}p^{n-k}q^{k}.
\label{g_Z_thermodynamic}
\end{eqnarray}
This concludes the proof of the theorem  in the thermodynamic limit
$L\rightarrow\infty$. For finite $L$ the expression of (\ref{g_z}) can be obtained
by solving a recurrence relation obtained from the analytical expressions
obtained from  (\ref{thermal_density_matrix_decomposition1}) for the cases $n=2,3,4$. The
correctness of this expression can be checked indirectly through the
expression for the eigenvalues (see \cite{EntanglementThermic}).

\section{Spectral properties of RDM}

Since RDM is block diagonalized with respect to quantum number $k$ into
blocks $B_{k}$ of size $\binom{n}{k}\times\binom{n}{k}$, to diagonalize the
complete RDM we can  diagonalize each block $B_{k}$, $k=0,1,...,n$
separately. As mentioned before, the permutational invariance
of the subsystem of size $n$  permits to further block diagonalize each  block $B_{k}$
with respect to a quantum number $s$ which labels the irreducible
representations of $\bf S_n$ (e.g. the YT of type $\{n-s,s\}$) which are compatible with the
block polarization. This implies that for a block $k$ the number $s$ can  take only
the values $s=0,...\min (k,n-k)$.
We  refer to the respective eigenvalues as $\lambda_{(s)}(n,k)$, omitting for
brevity the  explicit dependence on $L, N, r$. Note that the
eigenvalues $\lambda_{(s)}(n,k)$ have degeneracies $\binom{n}{s}-\binom{n}{s-1}$ which
are equal to the  dimension of the corresponding $\bf S_n$ representation (e.g.  the dimension of
the YT of type  $\{n-s,s\}$) \cite{EntanglementThermic}. These properties can be easily checked
on the particular example given in Fig. \ref{Fig_n6_twoblocks}).
By diagonalizing matrices $B_{k}$ for small $n$ one
finds that all eigenvalues $\lambda_{(s)}(n,k)$ are linear combinations of
the matrix elements $g_{Z}$ of the form:
\begin{equation}
\lambda_{(s)}(n,k)=\sum_{Z=0}^{\min(k,n-k)}\alpha_{Z}^{(s)}(n,k)g_{Z},
\label{eigenvalue_basis_coefficients}
\end{equation}
with  $\alpha_{Z}^{(s)}$, \textit{integer coefficients}, satisfying the
following properties:
\begin{eqnarray}
& &
\alpha_{0}^{(s)}(n,k)=1,   \label{alpha_0_property}
\\
& &
\sum_{Z=0}^{\min(k,n-k)}\alpha_{Z}^{(0)}(n,k)=\binom{n}{k},  \label{b2}
\\
&&
\sum_{Z=0}^{\min(k,n-k)}\alpha_{Z}^{(s)}(n,k) = 0\; \quad \text{for }s>0,  \label{b1}
\\
& &
\sum_{Z=0}^{\min(k,n-k)}\alpha_{Z}^{(s)}(n,k)\binom{Z}{p}=0,\;
p=0,1,...s-1,  \label{alpha_Z_property(d)}
\\
& & \sum_{Z=0}^{\min(k,n-k)}\alpha_{Z}^{(s)}(n,k)\binom{Z}{k}
=\alpha_{k}^{(s)}(n,k)= \\ && \qquad\qquad\qquad\qquad =(-1)^{s}\binom{n-k-s}{k-s}.
\label{alpha_Z_property(e)}
\end{eqnarray}

Note that the coefficients $\alpha$ do not depend on the characteristics of
the original state $L,N$ and $r$. The dependence of the RDM\ eigenvalues on
these parameters enters through the elements $g_{Z}$ (\ref{g_z}). This
implies that the above properties of $\alpha$ can be proved using special
cases. E.g. in case when all $g_{Z}=g_{0}$ , we are back to ground state
problem solved in \cite{Entanglement_Heisenberg}. Each block $B_{k}$ has
single nonzero nondegenerate eigenvalue $\lambda_{(0)}(n,k)=\binom{n}{k}%
g_{0} $, entailing (\ref{b1}),(\ref{b2}). \qquad

The property (\ref{alpha_0_property}) is easily proved in the thermodynamic
limit. In this limit, we have $\eta=0$. Then, all $g_{Z}\equiv0$, for $Z>0$
and all $B_{k}$ are diagonal implying $\lambda_{(s)}(n,k)=g_{0}$ for any $s$.
This gives (\ref{alpha_0_property}).

We have no proof for the properties (\ref{alpha_Z_property(d)}),(\ref%
{alpha_Z_property(e)}). Note that the (\ref{b1}) is a particular case of (%
\ref{alpha_Z_property(d)}) for $p=0$.

Given the exact form of the coefficients $\alpha_{Z}^{(s)}(n,k)$ for $Z=0,k$,
(\ref{alpha_0_property}) and (\ref{alpha_Z_property(e)}), the remaining
coefficients are obtained recursively from (\ref{alpha_Z_property(d)}).
E.g., the last eigenvalue of block $B_{k}$ with the degeneracy $\binom{n}{k}-%
\binom{n}{k-1}$ is obtained using recursively (\ref{alpha_Z_property(d)})
for $s=k$, and $p=s-1,s-2,..0$ which gives
\begin{equation}
\lambda_{(k)}(n,k)=\sum_{Z=0}^{k}(-1)^{Z}\binom{k}{Z}g_{Z}\text{, \ \ \ }
\alpha_{Z}^{(k)}(n,k)=(-1)^{Z}\binom{k}{Z}.  \label{last_eigenvalue}
\end{equation}
Note that the last eigenvalue does not depend on $n$, but only on $k$. Other
eigenvalues do depend on $n$ as well. For the last but one eigenvalue we
have
\beq
\alpha_{Z}^{(k-1)}(n,k)=(-1)^{Z}\left( \binom{k-1}{Z}-\binom{n-2k+1}{1}%
\binom{k-1}{Z-1}\right)
\label{last_but_one_eigenvalue}
\eeq
for $Z=0,1,..k$. Analogously one can check that
\begin{eqnarray}
\alpha_{Z}^{(k-2)}(n,k) =  (-1)^{Z} \left[
\binom{k-2}{Z}- \qquad\qquad\qquad\qquad\qquad\quad \right. \nonumber
\\ 2 \binom{n-2k+2}{1}
\binom{k-2}{Z-1}   \left.
 +   \binom{n-2k+2}{2} \binom{k-2}{Z-2}
\right], \nonumber \\
\alpha_{Z}^{(k-3)}(n,k) = (-1)^{Z} \left[ \binom{k-3}{Z} - 3 \binom{n-2k+3}{1}
\binom{k-3}{Z-1} \right. \nonumber  \\ \left.
 +   3 \binom{n-2k+3}{2} \binom{k-3}{Z-2} - \binom{n-2k+3}{3} \binom{k-3}{Z-3}
\right]. \nonumber
\end{eqnarray}
From these expressions  the existence of a recursive relation is evident, this
leading to the following general result
\beq
\alpha_{Z}^{(k-m)}(n,k)=(-1)^{Z}\sum_{i=0}^{m}(-1)^{i}\binom{m}{i}\binom{%
n-2k+m}{i}\binom{k-m}{Z-i}
\label{alpha_Z_general}
\eeq
with $m=0,1,...,k$. Using (\ref{alpha_Z_general}) together with
(\ref{eigenvalue_basis_coefficients}), (\ref{Gz_elements_thermodynamic_limit}),
(\ref{g_z}), one obtains the \textit{complete spectrum of the RDM}.

\section{Conclusion}

In summary, we have provided explicit analytical expression of  the
reduced density matrix  of a subsystem of  arbitrary size $n$ of a permutational
invariant quantum many body system of arbitrary size $L$ and  characterized by
a state of  arbitrary permutational symmetry. We have shown, on the specific example of the
spin $1/2$ Heisenberg model, that the RDM acquires a block diagonal form with respect to
the quantum number $k$ fixing the polarization in the subsystem  (conservation of $S_z$ or
conservation of the number of particles for non spin models )
and with respect to the irreducible representations of the $\bf S_n$ group.
The main results of the paper are represented by Eqs. (\ref{g_z}),
(\ref{Gz_elements_thermodynamic_limit}) and (\ref{alpha_Z_general})
given above. These results, being based only on the permutational invariance and on the conservation of
$S_z$ (number of particles for non spin systems),  should apply to all quantum many-body
systems which possess these symmetries and  for which the mean field theory becomes exact.
A more detailed analysis of the properties of
RDM spectrum and of its physical implications will be given elsewhere \cite{Entanglement_Theorems}.

\vskip 1 cm
{\bf Acknowledgments}
This work is dedicated to the memory of Prof. Alwyn Scott. MS
acknowledges financial support from a MIUR interuniversity
initiative. VP thanks the University of Salerno for a two year
research grant (Assegno di Ricerca no. 1508) and the
Department of Physics "E.R.Caianiello" for the hospitality.

\end{document}